\documentclass[pra,aps,twocolumn,groupedaddress,nofootinbib,superscriptaddress]{revtex4}
\usepackage{graphicx}
\usepackage[nointegrals]{wasysym} %
\usepackage[export]{adjustbox}
\usepackage{amsmath,amsfonts,amssymb,latexsym}
\usepackage{hhline}
\usepackage{bm}
\usepackage{verbatim}
\usepackage{enumitem}
\usepackage{mathrsfs}
\usepackage{slashed}
\usepackage{empheq}

\newcommand{\p}{\partial}

\newcommand{\lan}{\langle}
\newcommand{\ran}{\rangle}

\newcommand{\da}{{\dagger}}
\newcommand{\doa}{\downarrow}
\newcommand{\upa}{\uparrow}
\newcommand{\ob}[1]{\mkern 1.5mu\overline{\mkern-1.5mu#1\mkern-1.5mu}\mkern 1.5mu}

\newcommand{\ra}{\rightarrow}
\newcommand{\lra}{\leftrightarrow}

\newcommand{\wt}{\widetilde}

\renewcommand{\(}{\left(}
\renewcommand{\)}{\right)}
\renewcommand{\[}{\left[}
\renewcommand{\]}{\right]}
\newcommand{\mt}{\mapsto}

\newcommand{\tp}{\otimes}

\newcommand{\twp}{{2\pi}}

\newcommand{\D}{\nabla}

\newcommand\bpm            {\begin{pmatrix}}
	\newcommand\epm           {\end{pmatrix}}

\newcommand{\bs}{\bigskip}

\def\app#1#2{%
	\mathrel{%
		\setbox0=\hbox{$#1\sim$}%
		\setbox2=\hbox{%
			\rlap{\hbox{$#1\propto$}}%
			\lower1.1\ht0\box0%
		}%
		\raise0.25\ht2\box2%
	}%
}

\newcommand{\tw}{\textwidth}

\newcommand{\inv}{^{-1}}

\newcommand{\ope}\odot

\usepackage{manfnt}

\renewcommand{\prl}{{\, \parallel \,}}
\newcommand{\bi}{\begin{itemize}}
	\newcommand{\ei}{\end{itemize}}

\usepackage{amsthm}

\newtheorem{theorem}{Theorem}

\theoremstyle{definition}
\newtheorem{definition}{Definition}
\theoremstyle{definition}

\newcommand\bd            {\begin{definition}}
	\newcommand\ed            {\end{definition}}
\newcommand\bt            {\begin{theorem}}
	\newcommand\et            {\end{theorem}}
\newcommand\be            {\begin{equation}}
	\newcommand\ee            {\end{equation}}
\newcommand\ba            {\begin{aligned}}
	\newcommand\ea            {\end{aligned}}
\newcommand\bea{\begin{equation}\begin{aligned}}
		\newcommand\eea{\end{aligned}\end{equation}}

\usepackage{subcaption}

\newcommand{\ethan}[1]{ { \color{blue} \footnotesize \textsf{ethan: \textsl{#1}} }}

\usepackage{hyperref} 
\hypersetup{final}
\hypersetup{colorlinks, citecolor=red, linkcolor=red, urlcolor=red} 

\newcommand{\sss}{\subsubsection}
\renewcommand{\ss}{\subsection}

\renewcommand{\a}{\alpha}

\renewcommand{\d}{\delta}
\newcommand{\De}{\Delta}

\newcommand{\G}{\Gamma}
\newcommand{\s}{\sigma}

\renewcommand{\l}{\lambda}
\renewcommand{\L}{\Lambda}
\renewcommand{\t}{\theta}

\renewcommand{\o}{\omega}
\renewcommand{\O}{\Omega}

\renewcommand{\c}{\chi}

\newcommand{\bfK}{\mathbf{K}}

\newcommand{\bfS}{\mathbf{S}}

\newcommand{\bfa}{\mathbf{a}}

\newcommand{\bfq}{\mathbf{q}}
\newcommand{\bfr}{\mathbf{r}}

\newcommand{\bfv}{\mathbf{v}}

\newcommand{\bfy}{\mathbf{y}}

\newcommand{\qq}{\qquad}

\newcommand{\mco}{\mathcal{O}}

\newcommand{\mce}{\mathcal{E}}

\newcommand{\mcd}{\mathcal{D}}
\newcommand{\mcl}{\mathcal{L}}

\newcommand{\mct}{\mathcal{T}}

\newcommand{\mca}{\mathcal{A}}
\newcommand{\mcp}{\mathcal{P}}

\newcommand{\mcv}{\mathcal{V}}

\newcommand{\mci}{\mathcal{I}}

\newcommand{\sfV}{\mathsf{V}}

\newcommand{\sft}{\mathsf{t}}

\usepackage[mathscr]{eucal} %

\usepackage{braket}

\renewcommand{\k}{\ket}

\usepackage{dcolumn}
\usepackage{pifont}
\usepackage{ulem}

\usepackage{scalerel}
\begin{document}
	
	\title{Non-Fermi liquids from kinetic constraints in tilted optical lattices}
	\author{Ethan Lake}\email{elake@mit.edu}
	\affiliation{Department of Physics, Massachusetts Institute of Technology, Cambridge, MA, 02139}
	\author{T. Senthil}\email{senthil@mit.edu}
	\affiliation{Department of Physics, Massachusetts Institute of Technology, Cambridge, MA, 02139}

	\begin{abstract}
		
		We study Fermi-Hubbard models with kinetically constrained dynamics that conserves both total particle number and total center of mass, a situation that arises when interacting fermions are placed in strongly tilted optical lattices. Through a combination of analytics and numerics, we show how the kinetic constraints stabilize an exotic non-Fermi liquid phase described by fermions coupled to a gapless bosonic field, which in several respects mimics a dynamical gauge field. This offers a novel route towards the study of non-Fermi liquid phases in the precision environments afforded by ultracold atom platforms. 
		
	\end{abstract}
	
	\maketitle

	\paragraph*{Introduction:} 
	
	A major ongoing program in quantum many body physics is the characterization of phases of matter in which the quasiparticle paradigm breaks down. The most striking examples where this occurs are non-Fermi liquids (NFLs), believed to describe the observed strange metal behavior in a number of quantum materials. The low energy excitations in NFLs typically admit no quasiparticle-like description, with their ground states instead being more aptly thought of as a strongly-interacting quantum soup. Our understanding of such states of matter, as well as the conditions in which they may be expected to occur, is very much in its infancy. To this end, it is extremely valuable to have examples of simple microscopic models in which NFLs can be shown to arise, especially so when these models are amenable to experimental realization. 
	
	In this work we propose just such a model, by demonstrating the emergence of a NFL in a kinetically constrained 2d Fermi-Hubbard model. 
	This model is interesting in its own right, but our interest derives mainly from the fact that it finds a natural realization in strongly tilted optical lattices, a setup which has received recent experimental attention as a platform for studying ergodicity breaking and anomalous diffusion \cite{guardado2020subdiffusion,scherg2021observing,zahn2022formation}. 	
	The key physics afforded by the strong tilt is that it provides a way of obtaining dynamics that conserves both total particle number {\it and} total dipole moment (for us `dipole moment' is synonymous with `center of mass'), with the latter conserved over a prethermal timescale which as we will see can be made extremely long. 
	
	In different settings, the kinetic constraints provided by dipole conservation are well-known to arrest thermalization and produce a variety of interesting dynamical phenomena \cite{sala2020ergodicity,khemani2020localization,rakovszky2020statistical,van2019bloch,taylor2020experimental}. More recently, it has been realized that dipole conservation also has profound consequences for the nature of quantum ground states \cite{lake2022dipolar,lake2022dipole,zechmann2022fractonic,yuan2020fractonic,chen2021fractonic,sachdev2002mott,sachdev2011quantum} and the patterns of symmetry breaking that occur therein \cite{stahl2021spontaneous,kapustin2022hohenberg}. 
	
	In this work, we show that when these constraints arise in the context of Fermi-Hubbard models, they produce an exotic NFL state in an experimentally-accessible region of parameter space. The low energy theory of this NFL is closely analogous to a famous model in condensed matter physics, namely that of a Fermi surface coupled to a dynamical $U(1)$ gauge field \cite{holstein1973haas,halperin1993theory,lee1992gauge,lee2006doping,lee2018recent}. 
	We leverage this analogy to derive a number of striking features of the NFL state, chief among these being the absence of quasiparticles despite the presence of a sharp Fermi surface, and a vanishing conductivity despite the presence of a nonzero compressibility.

	\paragraph*{Fermions in strongly tilted optical lattices: } 
	
	We begin by considering a model of spinless fermions on a tilted 2d square optical lattice, interacting through repulsive nearest-neighbor interactions (spinful fermions are similar and will be briefly discussed later). Writing the fermion annihilation operators as $f_\bfr$ and letting $n_\bfr \equiv f^\da_\bfr f_\bfr$, we consider the microscopic Hamiltonian $H = H_{FH} + H_{\De}$, with the lattice tilt captured by $H_{\De} = \sum_{\bfr,\bfa} \De_a r^a n_\bfr$, and the Fermi-Hubbard part given by 
	\bea H_{FH} & = \sum_{\bfr, \bfa} [-t_a (f^\da_\bfr f_{\bfr+\bfa} +h.c.) +V_{0,a} n_\bfr n_{\bfr+\bfa}],
	\eea 
	where $\bfa = \hat{x}, \hat{y}$ label the unit vectors of the square lattice. The bare nearest-neighbor repulsion $V_{0,a}$ can be engineered by employing atoms with strong dipolar interactions \cite{baier2016extended,phelps2020sub} or by using Rydberg dressing \cite{pupillo2010strongly,johnson2010interactions,guardado2021quench}. To simplify the notation we will let both $t_a / \De_a$ and $V_0 \equiv V_{0,a}$ be independent of $a$, with $\De_x / \De_y = t_x / t_y$ unconstrained.
	
	We will be interested in the large-tilt regime $t_a/\De_a, V_0 / \De_a \ll 1$, with $t_a / V_{0}$ arbitrary. Here it is helpful to pass to a rotating frame which eliminates $H_\De$ via the time-dependent gauge transformation $e^{i t H_\De}$. In this frame, the Hamiltonian is 
	\be \label{hrot} H_{rot}(t) = \sum_{\bfr,\bfa} [-t_a(e^{-i\De_a t} f^\da_\bfr f_{\bfr+\bfa} + h.c.) + V_{0} n_\bfr n_{\bfr+\bfa}].\ee 
	We then perform a standard high-frequency expansion \cite{goldman2014periodically,eckardt2015high,bukov2015universal,mikami2016brillouin,scherg2021observing,mori2022heating} to perturbatively remove the quickly oscillating phases in the first term. 
	The time-independent part of the resulting expansion conserves the total dipole moments $\sum_\bfr r^a n_\bfr$ because dipoles---being charge neutral objects---can hop freely without picking up any $e^{-i\De_a t}$ phases. 
	As shown in App. \ref{app:eff_hams}, the result of this expansion is the static Hamiltonian 
	\bea \label{dfhm} & H_{DFH} = - \sum_{\bfr,\bfa} t_d[ d^{a\da}_\bfr( d^a_{\bfr + 2\bfa} + d^{a}_{\bfr+\bfa + \ob \bfa } + d^{a}_{\bfr+\bfa - \ob\bfa}  + h.c.)] \\ 
	& + \sum_{\bfr,\bfa}n_\bfr( V n_{\bfr+\bfa} + V'  n_{\bfr+2\bfa} + V''(n_{\bfr+\bfa + \ob\bfa} + n_{\bfr + \bfa - \ob\bfa}))\eea 
	where we have defined the dipole operators $d_\bfr^a \equiv f^\da_\bfr f_{\bfr+\bfa}$ and let $\ob a$ be the spatial coordinate opposite to $a$. The couplings constants in $H_{DFH}$ are given by
	\be \label{dfhm_coeffs} t_d = V_0 \sft^2, \, V = V_0(1-6\sft^2), \, V' = 2t_d,\, V'' = 4t_d,\ee
	with the dimensionless hopping strength $\sft \equiv t_a / \De_a$.  
	We will refer to the model \eqref{dfhm} as the {\it dipolar Fermi-Hubbard model} (DFHM). 
	
	As a time-independent theory, the DFHM only captures the system's dynamics over a (long) prethermal timescale. For (yet longer) times the fermions can exchange energy between $H_{FH}$ and $H_{\De}$, and a system initially prepared in the ground state of $H_{DFH}$ will begin to heat up.
	We will see later that this is actually not an issue, as the relevant time scale can (in principle) be made arbitrarily long. Before explaining this however, we first turn our attention to understanding the low-energy physics of $H_{DFH}$.
	
	\begin{figure}
		\vstretch{1.07}{\includegraphics[width=.5\tw]{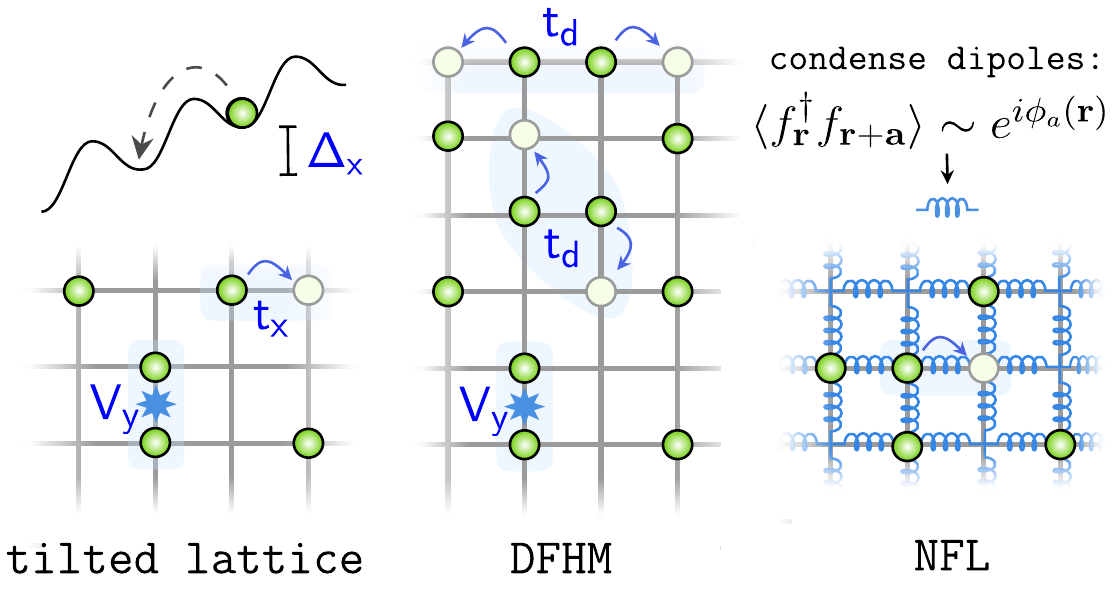}}
		\caption{\label{fig:insp} {\it Tilted lattice}: We consider an extended Hubbard model in a tilted optical lattice, with single particle hoppings $t_{x,y}$, nearest neighbor repulsions $V_{x,y}$, and tilts along both directions with strengths $\De_{x,y}$. {\it DFHM}: In the large tilt limit the system is described by a dipole-conserving Fermi-Hubbard model, whose dynamics is such that only dipolar bound states---rather than individual fermions---are allowed to move. {\it NFL:} As the dipole hopping strength $t_d$ increases the dipoles condense, with $ f^\da_\bfr f_{\bfr+\bfa} \sim e^{i\phi_a(\bfr)}$ developing an expectation value, and $\phi_a$ effectively playing the role of the spatial components of a dynamical $U(1)$ gauge field. The condensate liberates the motion of individual fermions, which form a Fermi surface. As described in the main text, in this regime fluctuations in $\phi_a$ turn the system into a non-Fermi liquid. } 
	\end{figure}

	\paragraph*{Theory of the dipolar Fermi-Hubbard model: }

	In the DFHM, dipole conservation fixes the center of mass of the fermions, which cannot change under time evolution. This precludes a net flow of particles in any many-body ground state, implying a particle number conductivity which is strictly zero at all frequencies and guaranteeing that $H_{DFH}$ always describes an insulating state \cite{lake2022dipolar}. In clean systems, a vanishing conductivity almost always comes hand-in-hand with a vanishing compressibility $dn/d\mu = 0$. We will however see that for a wide range of $\sft$ the natural ground state of the DFHM is in fact {\it compressible}. In this regime the system has a sharp Fermi surface but lacks well-defined Landau quasiparticles, and is therefore an example of a NFL.
	
	To understand the claims in the previous paragraph, we start by considering the limit of small $\sft$. Here the repulsive interactions dominate, and various crystalline states may form in a manner dependent on the fermion density. As $\sft$ is increased, the system can lower its energy by letting dipolar bound states delocalize across the system, by virtue of the dipole hoppings on the first line of \eqref{dfhm}. 
	For large enough $\sft$ the dipoles will lower their kinetic energy by condensing, producing a phase where $D^a \equiv \lan d^a\ran \neq 0$ and spontaneously breaking the dipole symmetry. When applied to the Hamiltonian \eqref{dfhm} at half-filling, a mean-field treatment (see App. \ref{app:mf}) predicts a condensation transition $\sft =1/4$, a value small enough that the perturbative analysis leading to $H_{DFH}$ should remain qualitatively correct.
	
	For simplicity we consider an isotropic dipole condensate, with $|D^x| = |D^y| \equiv D$. As also happens in the bosonic version of this model \cite{lake2022dipolar,lake2022dipole,zechmann2022fractonic}, the dipole condensate liberates the motion of single fermions, as they can move by absorbing dipoles from the condensate. Writing $D^a \simeq De^{i \phi_a}$ and taking $D$ to be constant (as amplitude fluctuations are gapped in the condensate), the first line of \eqref{dfhm} becomes the {\it single}-fermion hopping term
	\be
	\label{hhop}	H_{hop} = -t_dD\sum_{\bfr,\bfa}  ( f^\dagger_{\bfr} e^{i\phi_a(\bfr)} f_{\bfr + \bfa} + h.c. ).
	\ee 
	
	If we freeze out the dynamics of $\phi_a$, $H_{hop}$ will lead the fermions to form a Fermi surface, with an area set by their density as per Luttinger's theorem. 
	The important question is then to ask what happens when one accounts for the dynamics of the $\phi_a$ fields. As soon as we introduce these dynamics, the system looses its ability to respond to uniform electric fields, and is rendered insulating. 
	Indeed, turning on a background vector potential $A_a$ in $H_{hop}$ simply amounts to replacing $\phi_a$ by $\phi_a + A_a$. We can then completely eliminate the coupling of the fermions to $A_a$ through a shift of $\phi_a$ (see also \cite{shi2022loop}). Since $\phi_a$ is the Goldstone mode for the broken dipole symmetry, all other terms in the effective Hamiltonian can only involve gradients of $\phi_a$, and thus after the shift, the Hamiltonian can only depend on gradients of $A_a$. This then leads to a particle conductivity $\s(\o,\bfq)$ that vanishes for all $\o$ as $\bfq \ra 0$. 
	
	To deepen our understanding of this phase, we pass to a field theory description by writing $f_{\bfr} \simeq \int d\theta \, e^{i \bfK_F(\theta) \cdot \bfr} \psi_\theta(\bfr)$,	where $\bfK_F(\theta)$ is the Fermi momentum at an angle $\theta$ on the Fermi surface. 
	Standard arguments then lead to the imaginary-time Lagrangian 
	\bea \label{s} 
	\mcl_{DFH} & = \int d\t \,  \psi^\da_\t \Big(\p_\tau - i\bfv_\t \cdot \D + \frac{\varkappa}2 \D_\prl^2  + \sum_a g_a(\t) \phi_a \Big) \psi_\t \\ &  \qq + \sum_a \kappa_D (\p_\tau \phi_a)^2 +  \sum_{a,b} K_{a,b} (\D_a \phi_b)^2.  \eea 
	%
	%
	In writing the above we have approximated the dispersion of $\psi_\t$ to include only the leading terms in the momentum deviation from $\bfK_F(\t)$, written the Fermi velocity as $\bfv_\theta$, let $\varkappa$ denote the Fermi surface curvature, and taken $\D_\prl$ as the derivative along the Fermi surface. 
	
	In the important `Yukawa' term $g_a(\t) \phi_a \psi^\da_\t \psi_\t$, the coupling function $g_a(\t)$ is strongly constrained by dipole symmetry, which sends $\psi_\theta \ra e^{i {\bf \alpha} \cdot \bfr} \psi_{\theta}$, $\phi_a \ra \phi_a + { \alpha}_a$ for any constant vector $\a_a$. The requirement that \eqref{s} be invariant then gives the constraint
	\begin{equation}
		\label{eq:gtov}
		g_a(\theta) = - v_a(\theta). 
	\end{equation}
	This implies that $\phi_a$ couples to the fermions in {\it exactly} the same way as the spatial part of a $U(1)$ gauge field. This draws a connection between the DFHM and a Fermi surface coupled to a dynamical $U(1)$ gauge field, a system with a long history in condensed matter physics. In both models the modes that couple to the fermions are vector fields that are guaranteed to be gapless---by gauge invariance in the gauge field case, and by their origin as Goldstone modes in the DFHM. Crucially, the coupling between $\phi_a$ and the fermions is not ``soft", remaining nonzero even at zero momentum (soft couplings are irrelevant under RG, and fail to induce NFL behavior). In line with the general framework of Ref. \cite{watanabe2014criterion}, this is made possible by the fact that the dipole charge and the total momentum $P^b$ satisfy $\langle [\sum_\bfr r^a n_\bfr , P^b] \rangle = i\delta^{a,b} \sum_\bfr \langle n_\bfr \rangle \neq 0,$ the non-vanishing of which is necessary to avoid obtaining a soft coupling.

	An important difference compared to the Fermi surface + gauge field problem is that in the DFHM, there is no analog of a time component of the gauge field. For fermions coupled to a dynamical gauge field $a_\mu$, the coupling to $a_0$ renders the theory incompressible: an external probe potential $A_0$ (the susceptibility to which the compressibility corresponds) evokes no response, as $A_0$ can be absorbed into $a_0$. Since there is no analogue of $a_0$ in the DFHM this argument does not apply, and indeed it is well-known that a Fermi surface coupled to a gapless boson is generically compressible \cite{lee2006doping,metlitski2010quantum,mross2010controlled,chubukov2018fermi,shi2022gifts}. Remarkably, we thus manage to obtain a system with both vanishing conductivity {\it and} nonzero compressibility (as also occurs in the ``Bose-Einstein insulator" phase of the dipole-conserving Bose-Hubbard model \cite{lake2022dipolar}).  
	
	To demonstrate the NFL nature of $\mcl_{DFH}$, we note that it is essentially the 
	same as the action that arises at the `Hertz-Millis' theory \cite{hertz1976quantum,millis1993effect} of the quantum critical point associated with the onset of loop current order in a metal \cite{shi2022loop} (but with the crucial restriction \eqref{eq:gtov} coming from dipole conservation). Like in that case, fluctuations of $\phi_a$ turn the system into an NFL. Indeed, standard calculations show that at the Fermi surface, the fermion self energy has the form $\Sigma_f(\bfK, i\omega) = i\, {\rm sgn}(\omega) |\omega|^\delta$
	with the exponent $\delta < 1$ (a variety of theoretical approximations all converge on $\delta = 2/3$ \cite{halperin1993theory, lee2006doping,lee2009low,metlitski2010quantum,mross2010controlled,esterlis2021large}, which is also the exponent of the low-temperature specific heat, $C \sim T^\d$). 
	This shows that there are no sharply-defined quasiparticles in this model, despite the existence of a sharply-defined Fermi surface. We also note that this model has no weak-coupling pairing instability, due to the strong repulsive interaction between fermions on antipodal patches mediated by $\phi_a$ \cite{metlitski2015cooper}. 
	

	\begin{figure}
		\includegraphics[width=.24\tw]{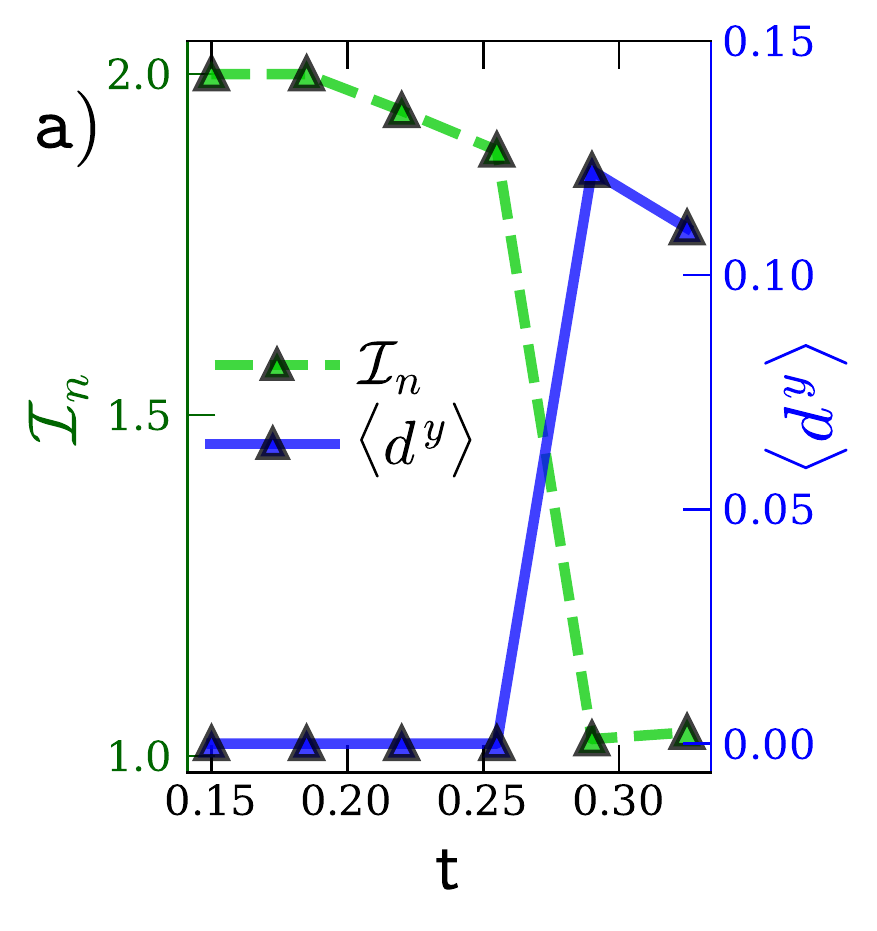}~\hfill\vstretch{.98}{\includegraphics[width=.215\tw]{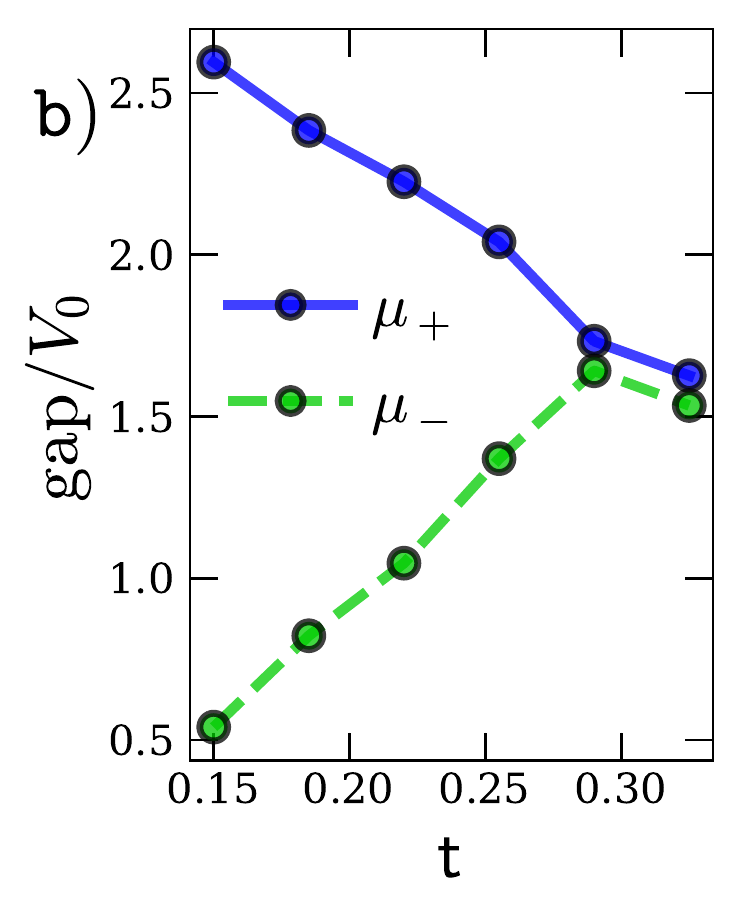}}
		\caption{\label{fig:dmrg} DMRG results for the dipolar Fermi-Hubbard Hamiltonian $H_{DFH}$ at half-filling on a cylinder of size $(L_x,L_y) = (20,6)$ and at bond dimension $\c = 400$. \texttt{a)} the inverse participation ratio $\mci_n$ of the density and the expectation value $\lan d^y_\bfr \ran = \lan  f^\da_\bfr f_{\bfr+\bfy}\ran $ averaged over lattice sites. As judged by $\mci_n$ charge order occurs at small $\sft$ but melts at $\sft_* \approx 0.275$; the plot of $\lan d^y\ran$ shows this is also where dipole condensation occurs. \texttt{b)} The energy cost to add or remove a fermion, $\mu_\pm  \equiv \pm E(N\pm1) \mp E(N)$, with $E(N)$ the ground state energy in the sector with total charge $N$. The charge gap $\mu_+ - \mu_-$ closes at the same location where dipole condensation occurs, suggesting the onset of the NFL. } 
	\end{figure}
	
	\paragraph*{Numerics: } 
	
	We now provide a first step towards testing the above theoretical predictions by performing DMRG on small cylinders with the DFHM Hamiltonian \eqref{dfhm}. We focus on the case of half-filling so as to compare with predictions from mean field, which predicts a dipole condensation transition at $\sft = 1/4$. Fig. \ref{fig:dmrg} shows the DMRG results for a cylinder of modest size $(L_x,L_y) = (20,6)$. For a range of $\sft$ near $1/4$ we compute the expectation values of the dipole operators and the inverse participation ratio 
	$\mci_n \equiv L_xL_y \lan \sum_i n_i^2 \ran / (\sum_i \lan n_i\ran )^2$ (Fig. \ref{fig:dmrg} left; $\lan d^x\ran$ is similar to $\lan d^y\ran$ but smaller in magnitude, presumably due to finite-size effects). We find $\mci_n \approx 2$, $\lan d^y\ran = 0$ for $\sft \leq \sft_*$ (as expected from a charge-ordered state) and $\mci_n \approx 1$, $\lan d^y\ran > 0$ for $\sft > \sft_*$ (as expected from a dipole condensate), where the critical value $\sft_* \approx 0.275$ is respectably close to the mean-field estimate. 
	
	To investigate the state at $\sft > \sft_*$ we compute the chemical potentials $\mu_\pm \equiv \pm E(N\pm1) \mp E(N)$, where $E(N)$ is the ground state energy in the symmetry sector with total charge $N$. The gap to charged excitations is given by $\mu_+ - \mu_-$, which is seen to approximately close at $\sft_*$ (Fig. \ref{fig:dmrg}, right). This suggests that at $\sft_*$ the system undergoes a (presumably first-order) transition into the NFL described by \eqref{s}. While this is all in accordance with our theoretical analysis, these numerics do not answer questions about the doping dependence of $\sft_*$, or reveal the nature of the correlations present in the NFL phase. A proper treatment of these questions is left to future work.

	\paragraph*{Experimental considerations: }
	
	The most obvious experimental signature of the NFL state is the simultaneous presence of both a nonzero compressibility and a vanishing particle conductivity, both of which can be directly measured from density snapshots taken in quantum gas microscopes \cite{brown2019bad,hartke2020doublon}. The dipole condensate can also be directly detected by measuring \cite{kessler2014single,atala2014observation} correlation functions of the dipole operators $f^\da_\bfr f_{\bfr+\bfa}$; these correlation functions are long-ranged in the NFL but short-ranged in the charge-ordered state. 
	The Fermi surface itself can be detected in principle by looking for Friedel oscillations in the density-density correlation function \cite{riechers2017detecting}, which are even stronger \cite{mross2010controlled} than in a conventional Fermi liquid. 
	
	We now discuss issues relating to the experimental preparation of the NFL state.
	In one possible protocol, the system is prepared in a uniform density product state at zero single particle hopping $t_a = 0$ and zero tilt $\De_a = 0$. $\De_a$ is then diabatically switched on to a value much larger than the Hubbard interaction and the dimensionless hopping strength $\sft \equiv t_a / \De_a$ is slowly increased, with the goal of reaching the NFL regime while keeping the dipole-conserving system at an effective temperature $T \lesssim T_F$, with $T_F \sim t_d = V_0 \sft^2$ the Fermi temperature. 
	
	At this point in the discussion, the prethermal nature of $H_{DFH}$ becomes important.  
	In going from \eqref{hrot} to \eqref{dfhm} we only kept the time-independent part of the effective Hamiltonian; a more complete analysis shows that in fact $H = H_{DFH} + \mcv(t)$, with the most important part of $\mcv(t)$ being $\mcv(t) = V_0 \sft^2 \sum_{\bfr;s=\pm1} e^{i(\De_x + s\De_y)t} \mco^{s}_\bfr + h.c,$ 
	where $\mco_\bfr^s$ is a rather complicated four-fermion interaction with a net dipole moment of 1 ($s$) in the $x$ ($y$) direction (see App. \ref{app:eff_hams} for the full expression). $\mcv(t)$ causes a system initially prepared in the ground state of $H_{DFH}$ to heat up. Furthermore, if $|\De_x | = |\De_y|$, $\mcv(t)$ contains time-{\it independent} terms which break one linear combination of the two components of the dipole moment symmetry (if $|\De_x| / |\De_y| = p/q$ is rational, such terms will arise at $q$th order in perturbation theory). Breaking the symmetry in this way will generically yield a nonzero conductivity along one spatial direction and produce a crossover to an anisotropic phase that preempts the NFL at large scales. Fortunately, we now argue that these problems are not as severe as they might appear.
	
	The issue of $\mcv(t)$ containing time-independent dipole-violating terms can be circumvented simply by taking $|\De_x | / |\De_y|$ to be irrational \cite{khemani2020localization}. However even when $|\De_x |  = |\De_y|$, the time-independent part of $\mcv(t)$ is highly irrelevant and only produces a violation of \eqref{eq:gtov} through 2-loop diagrams that are suppressed by further powers of $\sft$. In practice, these symmetry-breaking terms may thus only lead to a crossover out of the NFL at length scales larger than experimentally relevant system sizes. 
	
	To assess the effects of heating, we estimate the heating rate $r$ of a state initially prepared in the ground state of $H_{DFH}$ and then time-evolved with $H_{DFH} + \mcv(t)$. $r$ can be bounded using the theory of Floquet prethermalization \cite{abanin2015exponentially,kuwahara2016floquet,abanin2017rigorous,else2020long} as 
	\be \label{r} r <  C' V_0^2 e^{C |\De_x| / J},\ee 
	where $C,C'$ are dimensionless constants depending on $|\De_x| / |\De_y|$, and where $J$ is an energy scale determined by the maximum amount of energy locally absorbable by $H_{DFH}$ (if $|\De_x| / |\De_y|$ is irrational, $|\De_x|/J$ in the exponent is replaced by $\sqrt{|\De_x| / J}$ \cite{else2020long}).
	
	From the couplings given in \eqref{dfhm_coeffs}, we see that {\it all} of the terms in $H_{DFH}$ are proportional to $V_0$, meaning that $J = C''V_0$ for another dimensionless constant $C''$. Crucially though, the parameter $\sft$ which tunes between the different phases of $H_{DFH}$ is {\it independent} of $V_0$. This implies that by decreasing $V_0$ and keeping $\sft$ fixed we can make $|\De_x| / J$ arbitrarily large---and hence $r$ arbitrarily small---all while remaining at a {\it fixed} point in the phase diagram. 
	This parametric suppression of $r$ means that the issue of prethermal heating can in principle be sidestepped simply by working at weak bare interactions. 
	
	\paragraph*{Discussion:}

	We have demonstrated the emergence of a rather exotic non-Fermi liquid (NFL) from a simple dipole-conserving Fermi-Hubbard model. 
	This model has a natural realization in strongly tilted optical lattices, and in the NFL regime is described by fermions coupled to an emergent bosonic mode which plays the role of a spatial gauge field. 
	This provides an ultracold-atoms path towards the study of strongly interacting fermions and gauge fields in a manner rather distinct from approaches that build in gauge fields at a more microscopic level \cite{zohar2015quantum,aidelsburger2022cold,yang2020observation}.
	
	As always with ultracold atoms, the experimental crux is likely to be whether or not one can access the temperatures $T \lesssim T_F$ required to probe the physics of the NFL ground state. With this in mind it is natural to wonder if the kinetic constraints imposed by dipole conservation lead to any interesting {\it dynamical} signatures of the NFL regime, which could be more readily identified in experiments unable to perform a sufficiently adiabatic parameter sweep.
	
	While our focus so far has been on systems of spinless fermions, similar physics is also realizable in spinful models. A natural place to look is the tilted Fermi-Hubbard model
	\be H = \sum_{\bfr,\bfa,\s}[-t_a( f^\da_{\bfr,\s} f_{\bfr+\bfa,\s} + h.c. )+\De_a r^an_{\bfr,\s}] + V_0\sum_\bfr n_{\bfr,\upa} n_{\bfr,\doa}. \ee 
	In the large $\De_a$ limit, this Hamiltonian yields a dipole-conserving model with nearest-neighbor interactions and a Heisenberg exchange proportional to $V_0$. 
	At half filling and with {\it attractive} bare interactions, mean-field calculations predict a transition at $t_a / \De_a \gtrsim 0.26$, where the dipole operators $f^\da_{\bfr,\s} f_{\bfr+\bfa,\s}$ condense and subsequently produce an NFL. We leave a more thorough investigation of this physics to future work.


	\paragraph*{Ackgnowledgements:} We thank Monika Aidelsburger, Zhen Bi, Soonwon Choi, and Byungmin Kang for discussions, and Jung-Hoon Han, Hyun-Yong Lee, and Michael Hermele for collaborations on related work. T.S. was supported by US Department of Energy Grant No. DE- SC0008739, and partially through a Simons Investigator Award from the Simons Foundation. This work was also partly supported by the Simons Collaboration on Ultra-Quantum Matter, which is a grant from the Simons Foundation (Grant No. 651446, T.S. DMRG simulations were performed with the help of the Julia iTensor library \cite{fishman2022itensor}.
	
	\paragraph*{Note added: } While preparing this preprint we learned of a related and soon-to-appear work by A. Anakru and Z. Bi. 
	
	\bibliography{dfhm}
	
	\appendix

	\begin{widetext}
		
		\bs 
		
		\bs 
		
		\section{Realizing the dipolar Fermi-Hubbard model in strongly tilted optical lattices} \label{app:eff_hams}
		
		In this appendix we derive the effective dipole-conserving Hamiltonians governing the prethermal physics of fermions in strongly tilted optical lattices (see e.g. \cite{khemani2020localization,scherg2021observing,moudgalya2019thermalization,lake2022dipolar,lake2022dipole} for related calculations). We will do this using the van Vleck expansion \cite{bukov2015universal,eckardt2015high,mikami2016brillouin}, which also allows us to estimate the heating rate in the prethermal regime. 
		
		We will start with the simpler case of spinless fermions, with the results for spinful fermions quoted in a subsequent subsection. For notational simplicity we will suppress the boldface on vectors, writing e.g. $r+a$ for $\bfr+\bfa$. 
		
		\ss{Spinless fermions}
		
		Our starting point is the microscopic Hamiltonian associated with a tilted lattice of spinless fermions with tilt strengths $\De_a$, nearest-neighbor hoppings $t_a$, and bare Hubbard repulsions $V_{0,a}$: 
		\be\label{huv} H = \sum_{r;a=1,\dots,d} (r_a \De_a n_r -t_a(d^a_r + d^{a\da}_r) + V_{0,a} n_rn_{r+a} ) \equiv H_\De+H_t + H_V.\ee
		As in the main text, it is conceptually helpful to switch to a frame in which the tilt terms in the microscopic Hamiltonian are absent, while the single particle hopping terms acquire a rapidly oscillating time dependence. Now if the Schrodinger equation for $\k\psi$ reads $H \k\psi = i \p_t \k\psi$, then the Schrodinger equation for $\k{\psi_O} \equiv e^{iO(t)} \k\psi$ is governed by the rotated Hamiltonian 
		\be H_O = e^{iO(t)} H e^{-iO(t)} - e^{iO(t)} i\p_t e^{-i O(t)}.\ee 
		Thus starting from the time-independent tilted lattice Hamiltonian \eqref{huv}, we can rotate away the tilt term along the $a$ direction using the operator $\mcd_a  = t \sum_{r} \De_a r_a n_r$. The Hamiltonian $H_{\mcd_a}$ in the rotated frame is then 
		\bea H_{\mcd_a} (t) & = -\sum_{r} (t_a d^a_r e^{-i\De_a t}+h.c.) - \sum_{r,b\neq a} (t_b d^b_r + h.c.) + \sum_{r,b\neq a} \De_b r_b n_r +  H_V \\ 
		& \equiv T_a e^{-i\De_at} + h.c. + \ob{H_{\mcd_a}},\eea 
		where $\ob{H_{\mcd_a}}$ is the time-independent part of $H_{\mcd_a}(t)$ and we have defined the operator $T_a \equiv - \sum_r t_a d^a_r$.
		We now perform a second frame change via an operator $L^a(t)$ to remove the time dependence in the single particle hopping term $T_a e^{-i\De_a t} + h.c.$, now working perturbatively by taking $\De_a$ to be large compared with $t_a,V_{0,a}$ (we will fix $\De_a > 0 \ \ \forall \ \ a$ without loss of generality). After performing the frame change for the tilt along the $a$ direction we will come back and repeat the process for the other tilt directions, eventually eliminating all of the tilt terms. 
		

		In the van Vleck expansion, we write $L^a(t)$ and the resulting rotated Hamiltonian $H_{L^a}$ in a series expansion as
		\be L^a(t) = \sum_{n=1}^\infty \frac{L^a_k(t)}{\De_a^k},\qq H_{L^a} = \ob{H_{\mcd_a}} + \sum_{k=1}^\infty \frac{\O^a_k}{\De_a^k}.\ee 
		As explained in detail in Apps. B and C of \cite{mori2022heating}, the effective Hamiltonian up to a given order $n$ in the expansion is 
		\be H_{L^a,{\rm eff}}^{(n)} = \ob{H_{\mcd_a}} + \sum_{k=1}^n \frac{\O^a_k}{\De_a^k} + \frac1{\De^{n+1}_a} \p_t L^a_{n+1}(t).\ee 
		Using the techniques of \cite{mikami2016brillouin,mori2022heating}, it is straightforward to show that to the orders we will need them (third in $L^a$ and second in $\O^a$), we have\footnote{In the derivation of these equations our life is made considerably simpler by the fact that the time-dependent part of $H_{\mcd_a}(t)$ contains only a single Fourier harmonic.}
		\bea L^a_1(t) & = i (T_a e^{-i \De t} - h.c.)\\ 
		L^a_2(t) & = -i ([T_a ,\ob{H_{\mcd_a}}] e^{-i\De_a t}+ h.c.) \\ 
		L^a_3(t) & = i ([[T_a, \ob{H_{\mcd_a}}], \ob{H_{\mcd_a}}] e^{-i\De_at} - h.c.)\eea 
		and 
		\bea \O^a_1 & = 0 \\ 
		\O^a_2 & =\frac12 \( [[T_a,\ob{H_{\mcd_a}}],T_a^\da] + h.c.\).\eea 
		Truncating the expansion to this order in perturbation theory, the effective Hamiltonian is thus 
		\bea   H_{L^a,{\rm eff}}^{(2)} & = \ob{H_{\mcd_a}} + \frac1{2\De_a^2}\( [[T_a,\ob{H_{\mcd_a}}],T_a^\da] + h.c.\) + \frac1{\De_a^2} \( [[T_a, \ob{H_{\mcd_a}}], \ob{H_{\mcd_a}}] e^{-i\De_at} +h.c.\) . \eea 
		It is easy to check that $[T_a,T_b^{(\da)}] = [T_a, \mcd_b] = 0$ for all $a\neq b$ (at least up to boundary terms). This means we may write the above more explicitly as 
		\bea \label{hlaeff} H_{L^a,{\rm eff}}^{(2)} & = \sum_{b \neq a} \(T_b + T_b^\da + \sum_r \De_b r_b n_r\) + H_V + \frac1{\De_a^2} [[T_a,H_V],T_a^\da] +  \frac1{\De_a^2}  [[T_ae^{-i\De_a t} + T^\da_a e^{i\De_a t} ,H_V], \sum_{b\neq a} (T_b+T_b^\da) + H_V] \eea 
		Note that the time-independent part of $H^{(2)}_{L^a,{\rm eff}}$ conserves dipole moment along the $a$ direction, while the conservation is broken by the (last) time-dependent part, the appearance of which reflects the system's ability to exchange energy between the `tilt' and `non-tilt' parts of the microscopic Hamiltonian \eqref{huv}. 
		
		We have thus succeeded in eliminating the tilt in the $a$ direction. We proceed by eliminating the tilt in the remaining directions using a similar transformation, this time operating on $H^{(2)}_{L^a,{\rm eff}}$. Consider eliminating the tilt along the direction $c\neq a$. Rotating away the $c$ component of the tilt using the operator $\mcd_c$ results in removing $\sum_r \De_c r_c n_r$ from $H^{(2)}_{L^a,{\rm eff}}$ and replacing $T_c$ with $T_c e^{-i\De_c t}$. We then perform a van Vleck expansion with operators $L^c_k(t), \O_k^c$. 
		Since the last two terms in \eqref{hlaeff} are already at the highest order in perturbation theory that we are working to, they do not enter into the calculation of $L^c_k(t), \O_k^c$, simplifying the calculation. After successively eliminating all of the tilt directions, the resulting effective Hamiltonian (which we denote simply as $H_{\rm eff}$ to save space) is therefore 
		\bea \label{finalheff} H_{\rm eff} & = H_V + \sum_a \frac1{\De_a^2} [[T_a, H_V], T_a^\da] +  \sum_a \sum_{s=\pm1} \frac1{\De_a^2} [[T^s_a,H_V],H_V] e^{-is\De_a t} + \sum_{a\neq b} \sum_{s,s'=\pm1} \frac1{\De_a^2} [[T_a^s ,H_V],T_b^{s'}] e^{-i(s\De_a + s'\De_b)t} \\ 
		& \equiv H_{DFH} + \mcv(t), \eea 
		where $H_{DFH}$ ($\mcv(t)$) is the time-independent (time-dependent) part of $H_{\rm eff}$, and 
		where $T^s_a$ is defined to equal $T_a$ if $s=1$ and equal $T^\da_a$ if $s=-1$. 
		
		To proceed we need to compute the commutators appearing in \eqref{finalheff}. We start with 
		\bea  \label{twithv}[T_a ,H_V] & = \sum_{b,r} t_a V_{0,b} \((d^a_{r-a-b} -d^a_{r-b})n_r + n_r(d^a_{r-a+b} - d^a_{r+b})\)
		\\ &= 		\sum_{r,b} t_a V_{0,b} M_{a,b;r} d^a_r,  \eea
		where we have defined 
		\be M_{a,b;r} \equiv \d_{a,b}(n_{r+2a} - n_{r-a}) + (1-\d_{a,b})(n_{r+a+b} + n_{r+a-b} - n_{r+b} - n_{r-b}).\ee 
		
		\sss{$H_{DFH}$}
		
		To find $H_{DFH}$, we just need to take a commutator of \eqref{twithv} with $T_a^\da$. After some straightforward algebra, one finds  
		\bea\,\, [[T_a,H_V],T_a^\da] & = t_a^2\Big( V_{0,a} (-2d^{a\da}_{r-a}d^a_{r+a} + 2n_r(n_{r+2a} -n_{r+a}))  \\ 
		& + \sum_{b\neq a} V_{0,b} \(4d^{a\da}_r d^a_{r+b} - d^{a\da}_r \sum_{s,s'=\pm1}d^a_{r+sa+s'b}   + n_r \sum_{s,s'=\pm1} n_{r+sa+s'b} -  4n_r n_{r+b}  \). \eea 
		We thus may write  $H_{DFH}$ as 
		\bea \label{dfhm_app} H_{DFH} & = -\sum_{r,a} t_{d,a} (d^{a\da}_{r}d^a_{r+2a} + h.c.) -  \sum_{r,a\neq b}  [t_{d,ab}'(d^{a\da}_{r} d^a_{r+b} + h.c) + t_{d,ab}(d^{a\da}_r d^a_{r+a+b} + d^{a\da}_r d^a_{r+a-b} + h.c.)] \\ & + \sum_{r,a} \(V_a n_r n_{r+a} +  V_a' n_r n_{r+2a}\) + \sum_{a< b} V_{ab} n_r (n_{r+a+b} + n_{r-a+b}),\eea 
		where the various coupling constants appearing in the above are defined as 
		\bea t_{d,a} = \frac{t_a^2 V_{0,a}}{\De_a^2} \qq 
		t_{d,ab} = \frac{t_a^2 V_{0,b}}{\De_a^2} \qq 
		t_{d,ab}'  =- 2\frac{t_a^2 V_{0,b}}{\De_a^2} \eea 
		and 
		\bea 
		V_a  = V_{0,a}\(1 - 2 \frac{t_a^2 }{\De_a^2} - 4\sum_{b\neq a} \frac{t_b^2}{\De_b^2}\) \qq 
		V_{a}' = 2\frac{t_a^2 V_{0,a} }{\De_a^2} \qq 
		V_{ab} & =2(t_{d,ab} + t_{d,ba}).\eea 
		Note that as desired, $H_{DFH}$ conserves every component of the dipole moment. Also note that for all $a\neq b$, the dipole operators satisfy 
		\be  \label{dipconst} d^{a\da}_{r+b} d^a_r = - d^{b\da}_{r+a} d^b_r,\ee 
		which comes from the ambiguity present in grouping $f_r^\da f_{r+a} f^\da_{r+a+b} f_{r+b}$ either into two $a$-type dipoles or two $b$-type dipoles. For bosonic models the sign in \eqref{dipconst} is absent, and this relation is innocuous. In the current fermion case, the sign means that if $t'_{d,ab} = t'_{d,ba}$, the sum $\sum_{a\neq b} t_{d,ab}' (d^{a\da}_r d^a_{r+b} + h.c) = 0$, and transverse dipole motion is only possible by way of the `diagonal' hopping processes afforded by the $t_{d,ab}$ term.
		
		Finally, we quote what happens when a spatially-varying single-particle potential $H_\Gamma = \sum_r \Gamma_r n_r$ is added to the microscopic Hamiltonian. $H_\Gamma$ turns out to have a rather innocuous effect, with the only consequence being that $H_{DFH}$ then possesses the additional term 
		\be H_{DFH} \supset \sum_r n_r \( \G_r + \sum_a \frac{t_a^2}{\De_a^2} (\G_{r+a} + \G_{r-a} - 2\G_r) \),\ee
		so that perturbation theory simply renormalizes $\G_r$ by $\sum_a\partial_a^2 \G_r$. 
		
		\sss{$\mcv(t)$}
		
		We now derive the time-dependent part of the effective Hamiltonian \eqref{finalheff}. The first piece of desiderata is the commutator 
		\bea [[T_a,H_V],H_V] & = t_a  \sum_{r} \( \sum_b V_{0,b} M_{a,b;r}\)^2 d_r^a.\eea 
		The remainder of $\mcv(t)$ is determined by the commutators $[[T_a^s , H_V], T_b^{s'}]$ for $a\neq b$ (note that this is symmetric in $a,b$ since $[T^s_a , T^{s'}_b] = 0$). These are derived through unilluminating algebra, whose details we will omit. To simplify the resulting expression for $\mcv(t)$ we will specialize to two dimensions (the case we are most interested in), and for simplicity will set $V_{0,x} = V_{0,y} \equiv V_0$. Then 
		\bea \label{mcv} \mcv(t) & = \sum_{a,r} \frac{t_a V_0^2}{\De_a^2}\[ \( \sum_b M_{a,b;r} \)^2 d^a_r e^{-i\De_a t} + h.c.\] \\ 
		& \qq\qq+  e^{i(\De_x + \De_y)t}  t_x t_y V_0 \(\frac1{\De_x^2} + \frac1{\De_y^2} \) \Big[(f^\da_r f_{r+x+y} - f^\da_{r-y} f_{r+x})(n_{r+2x} - n_{r+x} + n_{r+x+y} - n_{r+y}) \\ 
		& \qq \qq + (n_r - n_{r-x} + n_{r+x-y} - n_{r-y})(f^\da_r f_{r+x+y} - f^\da_{r-y} f_{r+x}) \\ 
		&\qq\qq + \((d^x_{r-x} -d^x_r)(d^y_{r+x} - d^y_{r-y+x} + d^y_{r+y} - d^y_r) + (x\lra y) \) \Big] \\  &\qq \qq + ({\rm rot.\, \, sym.}), \eea 
		where (rot. sym.) denotes the three terms obtained from rotating the expressions in the second, third, and fourth lines in the plane by angles of $\pi/2, \pi,$ and $3\pi/2$ (with $(\De_x, \De_y)$ rotating as a vector). 
		
		In the case where $|\De_x| = |\De_y|$---so that the effective drive is periodic, rather than quasiperiodic---there will exist terms in $\mcv(t)$ which contain no time dependence. Such terms will ultimately break dipole symmetry (or rather, will break one linear combination of the $x$- and $y$-direction dipole symmetries, but preserve the other), and therefore likely ruin the NFL character of the NFL phase at long distances. However, as we now explain their effect is not likely to be too severe, and may be negligible for realistic system sizes. 
		
		Consider for concreteness the case when $\De_x = - \De_y \equiv \De$. Writing $\mcv(t) = \ob{\mcv} + \d\mcv(t)$, with $\ob\mcv$ time-independent and $\d\mcv(t)$ containing terms that oscillate as $e^{\pm i \De t}, e^{\pm 2i\De t}$, some algebra yields 
		\bea \label{obmcv} \ob \mcv & = 2 \frac{t_xt_y V_0}{\De^2} \sum_r \( f_r^\da ( \p_y n_{r-x +y} - \p_y n_{r+2x+y} + (x\lra y) ) f_{r+x+y}  + [d^x_r (\p_y d^y_{r+2x} + \p_x d^y_{r+x+y}) + (x\lra y)] \) + h.c, \eea 
		where we have defined the lattice derivative $\p_a \mco_r \equiv \mco_r - \mco_{r-a}$. \eqref{obmcv} follows from \eqref{mcv} after normal-ordering the terms where multiple fermion operators act on the same site. 
		
		We now understand the effects of $\ob\mcv$ in the IR theory $\mcl_{DFH}$ (cf. \eqref{s}). It is easy to see that $\ob\mcv$ does not directly modify either the Fermi velocity $\bfv_\t$ or the $g_a(\t) \phi^a \psi_\t^\da \psi_t$ Yukawa coupling, and thus does not induce a nonzero conductivity at tree-level. Indeed, including the leading fluctuations about the mean-field solution and expanding to leading order in derivatives, the dipole hopping terms in $\ob\mcv$ are represented in the IR as 
		\be \sum_r [d^x_r (\p_y d^y_{r+2x} + \p_x d^y_{r+x+y}) + h.c. ] + (x\lra y)\ra \int d\t \, \(\p_y \phi^{[y} \psi^\da_\t \p^{x]} \psi_\t  -\p_x \phi^{[x} \psi^\da_\t \p^{y]} \psi_\t  + (x\lra y)\)   + \cdots,\ee 
		where the $\cdots$ are four-fermion interactions. The term in parenthesis on the RHS above is identically zero, and thus the dipole hopping terms only produce a $\phi_a\psi^\da_\t\psi_\t$ vertex which is suppressed by even more derivatives, and thus is very irrelevant. 
		
		Instead, the main effect of $\ob\mcv$ is to produce dipole-breaking four-fermion interactions that can renormalize $g_a(\t),v_a(\t)$ in a way such that $g_a(\t) + v_a(\t) \neq 0$, thereby producing a nonzero conductivity.\footnote{The resulting theory will however not simply be a conventional Fermi liquid, since one linear combination of the two dipole moment generators (the antisymmetric combination when $\De_x = - \De_y$) will still remain a symmetry, enforcing zero conductivity along one spatial direction.} In the IR theory, these dipole-breaking interactions will not appear simply as leading-order Landau parameters or BCS interactions, since both of these interactions preserve dipole moment. Instead, they will appear as interactions like $\int_{\t,\t'} h^{a,b}_{\t,\t'} \psi^\da_\t \p_a \psi_\t \psi^\da_{\t'} \p_b \psi_{\t'}$, which are suppresed by two powers of momentum and are therefore rather irrelevant. While these interactions can lead to distinct renormalizations of $g_a(\t),v_a(\t)$ at the two-loop level, such an effect will only lead to a small nonzero value of $|g_a(\t) + v_a(\t)|$. Thus in practice the crossover out of the NFL state will likely be rather weak, and perhaps even negligible for experimentally realistic system sizes.

		\ss{Spinful fermions}
		
		The case of spinful fermions can be treated similarly. The general form of the effective Hamiltonian remains as in \eqref{finalheff}, but with the addition of spin indices in the appropriate places: one simply replaces $T_a$ by $T_{a,\s} \equiv t_{a,\s} \sum_r d^a_{r,\s}$ where $d^a_{r,\s} \equiv f_{r,\s}^\da f_{r+a,\s}$; $\De_a$ by $\De_{a,\s}$; sums over $\s$, and rotates the tilt potential away separately for each spin component. While continuing to work with a nearest-neighbor interaction is possible, in the spinful case it is more natural to instead let 
		\be H_V = V_0 \sum_r n_{r,\upa} n_{r,\doa},\ee 
		and we will do so in the following. This then gives 
		\be [T_{a,\s},H_V] = V_0 \sum_r t_{a,\s} ( d^a_{r-a,\s} - d_{r,\s}^a) n_{r,\bar\s}\ee 
		where $\bar \s$ is the opposite spin to $\s$. This allows one to derive 
		\bea [[T_{a,\s},H_V],T_{a,\s}^\da] & = 2 V_0 \sum_{r} t_{a,\s}\Big( -2t_{a,\s}n_{r,\s} n_{r,\ob \s} + 2t_{a,\s} n_{r,\s} n_{r+a,\ob\s} - 2t_{a,\ob\s} f^\da_{r,\s} f_{r,\ob\s} f^\da_{r+a,\ob \s} f_{r+a,\s} \\ \qq & + t_{a,\ob\s} f_{r,\s} f^\da_{r+a,\s} f^\da_{r+a,\ob\s} f_{r+2a,\ob \s} + h.c.\Big),\eea 
		which then gives us the time-independent part of the effective Hamiltonian. 
		We consequently find 
		\be \label{heff_spinful_app} H_{DFH} =-\sum_{r,a,\s} t_{d,a}  (f^\da_{r+a,\s} f_{r,\s} f^\da_{r+a,\ob\s} f_{r+2a,\ob\s} + h.c.) + V\sum_r n_{r,\upa} n_{r,\doa} + \sum_{r,a} \( \frac12J_{xy,a} (S^+_r S^-_{r+a} + h.c.) + J_{z,a} S^z_r S^z_{r+a}+V_a' n_r n_{r+a}\),\ee 
		where the dipole hopping strength is 
		\be \label{tdspinfull} t_{d,a}  = \frac{t_{a,\upa} t_{a,\doa} V_0}2 \( \frac1{\De_{a,\upa}^2} + \frac1{\De_{a,\doa}^2}\) ,\ee 
		and in terms of the parameter 
		\be \label{wttdspinfull} \wt t_{d,a} \equiv\frac{V_0}2 \sum_\s \frac{t_{a,\s}^2 }{\De_{a,\s}^2}\ee  (which equals $t_{d,a}$ if the single-particle hopping is spin-independent), the remaining coupling constants are 
		\bea 
		\label{spinful_coeffs}
		V & = V_0 - 4 \sum_{a} t_{d,a} \qq
		V_a' = \wt t_{d,a} \qq 
		J_{z,a}  = -4\wt t_{d,a} \qq 
		J_{xy,a}  = -4t_{d,a}. \eea 
		As a sanity check, specializing to 1d and taking $\De_\upa = \De_\doa$ can be checked to yield the same Hamiltonian as given in \cite{scherg2021observing}. Note that for small bare hopping strengths, attractive bare interactions ($V_0 < 0$) produce an anti-ferromagnetic Heisenberg exchange, while repulsive bare interactions produce a ferromagnetic one. 
		
		The time-dependent part of the effective Hamiltonian can be found in a similar fashion as in the spinless case; we leave a derivation of the explicit expression as an exercise to the reader.

		\section{Mean field theory for the dipole condensate} \label{app:mf}
		
		This appendix is devoted to a mean-field analysis of the DFHM, with the focus being on identifying the location of the phase transition (if present) where dipoles condense. We will start with the simpler case of spinless fermions (where the dipole condensate is easier to realize), and deal with spinful fermions (where the situation is more complicated) in the subsequent subsection. We will focus on fermions on the square lattice at half-filling throughout, where ground states at weak hopping strengths are easy to understand and the theoretical analysis is simplest. Half-filling is also where dipoles are most likely to condense: a dipolar involves both a local excess and deficit of charge, and thus their formation will be suppressed at fillings which are too low or too high; the particle-hole symmetric case of half-filling is thus the most natural place for dipoles to condense. 
		
		\ss{Spinless fermions}
		
		At half filling, the ground states of the $t_{d,a} = 0$ limit of the DFHM model \eqref{dfhm_app} are $\pi$-momentum CDW states. Without loss of generality we will focus on the ground state where the fermions occupy the $A$ sublattice, and will denote this state by $\k{CDW}$. There is a natural instability towards a dipole condensate\footnote{This is for dimensions $d>1$; in one dimension things are slightly different, and as detailed in App. \ref{app:spinless1d} an essentially exact solution is possible.}  when the dipole hoppings $t_{d,a}$ are turned on; in the context of tilted lattices this instability is made even stronger by virtue of the fact that the extended Hubbard terms $U'_a, U_{ab}$---which energetically disfavor $\k{CDW}$---are themselves proportional to the $t_{d,a}$. 
		
		We thus look for a mean-field decoupling in the dipolar (`excitonic') channel $d^a_r \equiv f^\da_r f_{r+a}$. Defining fields $D^a_r$ such that $\lan D^a_r\ran = \lan d^a_r\ran$, we may decouple the dipolar hopping term and write the imaginary-time action of the DFHM as 
		\be \label{spinless_imags} S = \int d\tau \, \( \sum_r c^\da_r \p_\tau c_r + \sum_{r,r',a} \mca^a_{r,r'} \( D^{a\da}_r D^a_{r'} - d_r^{a\da} D^a_{r'} - D^{a\da}_r d^a_{r'}  \) + H_n\),\ee 
		where $H_n$ is the part of the DFHM Hamiltonian that only involves number operators, viz.
		\be  H_n =  \sum_{r,a} \(V_a n_r n_{r+a} +  V'_a n_r n_{r+2a}\) + \sum_{a\neq b} V_{ab} n_r (n_{r+a+b} + n_{r-a+b}),\ee
		and where  
		\be \mca^a_{r,r'} \equiv t_{d,a} (\d_{r',r+2a} + \d_{r',r-2a}) + \sum_{b\neq a} \[t'_{d,ab}(\d_{r',r+b} + \d_{r',r-b}) + t_{d,ab}(\d_{r',r+a+b} + \d_{r',r+a-b} + \d_{r',r-a+b} + \d_{r',r-a-b})\].\ee 
		We now integrate out the fermions. For our purposes we will not need to find the quartic part of the effective dipole action, and will only focus on the quadratic piece, which is
		\bea \label{spinless_seff} S_{\rm eff}^{(2)} & = -\frac12 \int d\tau_1\, d\tau_2\, \lan CDW| \mct \prod_{j=1,2} H_{dD}(\tau_j) |CDW\ran + \sum_a \int d\tau \, D^{a\da}\mca^a D^a \\ 
		& = \int \frac{d\o}\twp \sum_{r,r',a} D^{a\da}_{r}(\o) D^a_{r'}(\o)\Big( -\frac{1}2 \int d\tau \, e^{-i\o \tau}\sum_{r'',n}  \mca^a_{r''r} \mca^a_{r''r'} \( |\lan CDW| d^{a\da}_{r''} |n\ran |^2 + |\lan CDW | d^a_{r''} |n\ran |^2  \) e^{-|\tau|E_{n0}} +   \mca^a_{rr'}\Big)\\
		& = \int \frac{d\o}\twp \sum_{r,r',a} D^{a\da}_{r}(\o) D^a_{r'}(\o) \Big(  -\sum_{r'',n}  \mca_{r''r}^a \mca_{r''r'}^a  \( |\lan CDW| d^{a\da}_{r''} |n\ran |^2 + |\lan CDW | d^a_{r''} |n\ran |^2  \)  \frac{E_{n0}}{\o^2 + E_{n0}^2}  + \mca^a_{rr'}\Big),\eea 
		where $H_{dD}$ is the term which couples $d$ to $D$ in \eqref{spinless_imags}, $\sum_n$ denotes a sum over all eigenstates of $H_n$, and $E_{n0} \equiv E_n - E_0$. We now need to calculate these energy differences. 
		
		The matrix elements $|\lan CDW| d^{a\da}_{r''} |n\ran |^2$ and $|\lan CDW | d^a_{r''} |n\ran |^2$ are only nonzero if $r''$ is in the $A$ or $B$ sublattice, respectively. The values of $E_{n0}$ are however the same for all states $\k n$ where these matrix elements are nonzero. We write let $\mce_a$ denote the energy differences for these states, with a simple calculation giving 
		\be \label{energy_gaps} \mce_a = V_a + 2\sum_{b\neq a} V_b - 2 \sum_{b\neq c} V_{bc} - 2\sum_b V'_b .\ee 
		Since in the tilted lattice context the $V_{ab},V'_a$ are positive and increase with $t_{a}$, while $V_a$ is positive and decreases with $t_a/\De_a$, there is obviously some value of $t_a/\De_a$ beyond which $\mce_a$ becomes negative (which decreases with larger spatial dimension $d$ because the number of terms in the sum $\sum_{b \neq c}$ increases quadratically with $d$), meaning that a mean-field solution will always exist for $t_a/\De_a$ sufficiently large. The only question remaining is how large the $t_a/\De_a$ need to be to get a solution; since we are doing perturbation theory in these parameters, we will get a self-consistent solution only if the dipole condensate occurs for small $t_a / \De_a$.
		
		After shifting $D^a \mt [\mca^a]\inv D^a$ and performing a derivative expansion on the $D^a$ fields, we may write the quadratic part of the continuum effective action as 
		\be \label{eft} S^{(2)}_{\rm eff} = \int d\tau \, d^dx\, \sum_a D^{a\da}(\tau,r) \( - \kappa_D^a\p_\tau^2  -\sum_{b,c} K^a_{b} \nabla_b^2 + r^a\) D^a(\tau,r),\ee 
		where from \eqref{spinless_seff} and \eqref{energy_gaps} we extract 
		\bea \label{coeffs} \kappa_D^a & = \frac1{\mce_a^3}  \\ 
		K^a_{b} & = \frac1{\sft_{d,a}^2} \( \d_{a,b} (4t_{d,a} + 2 \sum_{c\neq a} t_{d,ab}) + (1-\d_{a,b}) (t'_{d,ab} + 2t_{d,ab})\) \\ 
		r^a & = \frac1{\sft_{d,a}} - \frac1{\mce_a},
		\eea 
		where we have defined 
		\be \sft_{d,a} \equiv 2 \( t_{d,a} + \sum_{b\neq a} (t'_{d,ab} + 2t_{d,ab}) \).\ee 
		
		A condensate of $a$ dipoles will form when $D^a$ picks up an expectation value, i.e. when $r^a < 0$. From \eqref{coeffs}, this is seen to occur when  $\mce_a < \sft_{d,a}$. Consider now the case where the DFHM arises from a strongly tilted optical lattice, and for simplicity take $t_a / \De_a \equiv \sft$ and the bare Hubbard repulsion $V_{0,a}$ to be independent of $a$ (as discussed around \eqref{dipconst}, in this case $t'_{d,ab}$ can effectively be set to zero). Some algebra gives $\mce_a/ V_0 = (2d-1 - \sft^2(16d^2 - 12 d + 2))$ and $\sft_{d,a} = V_0\sft^2 (4d-2)$, so that dipole condensation occurs when
		\be \label{sftcond}   \sft \geq \frac1{\sqrt{8d}},\ee
		which for $d=2$ gives a (perhaps slightly uncomfortably large) value of $\sft \geq 1/4$. 
		Note that when $d=1$, the first inequality of \eqref{sftcond} exactly agrees with the estimate \eqref{1destimate} obtained from the spin-1/2 mapping available in one dimension.

		\ss{Spinful fermions}
		
		We now turn to the case of spinful fermions. We will consider the dipole-conserving Hamiltonian 
		\be  \label{spinful_h_app} H = \sum_{r,a,\s} t_d (f_{r,\s} f^\da_{r+a,\s} f^\da_{r+a,\ob\s} f_{r+2a,\ob\s} + h.c.) + V \sum_r n_{r,\upa} n_{r,\doa} +  \sum_{r,a} (J \bfS_r \cdot \bfS_{r+a} + V' n_r n_{r+a}),\ee 
		which for simplicity's sake we have taken to be slightly less general than the effective Hamiltonian \eqref{heff_spinful_app} (as the various coupling constants above are direction-independent and $H$ retains full $SU(2)$ symmetry---since in practice the tilt strengths are usually spin-dependent \cite{scherg2021observing} this is perhaps a slight over-simplification). As in the spinless case, we restrict our attention to fermions on the square lattice at half filling.
		
		\sss{attractive $V$}
		
		Consider first attractive interactions, $V <0$. The most favorable arrangement when $V$ dominates over $t_d,J,V'$ is to have a period-2 density wave $\k{CDW} \equiv \k{0,\upa\doa}^{\tp L/2}$. Dipoles can be created on top of this state, and if we imagine tuning $t_d$ independently of the other parameters, melting of the CDW by the formation of a dipole condensate should occur for large enough $|t_d / V|$. Note that whether or not this happens for the particular parametrization of coupling constants \eqref{tdspinfull}, \eqref{wttdspinfull}, \eqref{spinful_coeffs} given to us by the tilted lattice model is not immediately obvious: when we increase the bare single-particle hopping strength relative to the tilt we not only increase $t_d$, but also increase the nearest-neighbor interaction $V'$, which is attractive for $V < 0$. If the nearest-neighbor attraction gives a bigger energy decrease than the dipolar hopping term, the system may instead prefer to phase separate into voids and regions of full filling---thus a more detailed analysis is required. 
		
		We can proceed with a field theory analysis along the same lines as in the spinless case. One difference is that due to the spin degree of freedom, we can give our dipoles either spin 0 or spin 1. We will use a $\pm$ sign to distinguish between the two possibilities, denoting the dipole fields that decouple the dipole hopping term as $D_{r\s}^{a\pm}$, which we design to satisfy
		\be \lan D^{a\pm}_{r\s} \ran = \lan d^{a\pm}_{r\s}\ran,\qq d^{a+}_{r\s} \equiv f^\da_{r,\s} f_{r+a,\s},\qq d^{a-}_{r\s} \equiv f^\da_{r,\s} f_{r+a,\bar\s}.\ee 
		
		In terms of $D^s$, we may write 
		\be\label{sdecoup} S = \int d\tau \( f^\da \p_\tau f - s\,{\rm sgn}(t_d)\sum_a   (d^{s\da}_a\wt\mca^a D^{as} + h.c.) +s\,{\rm sgn}(t_d) \sum_a D^{as\da}  \wt\mca^a D^{as} + H_n\),\ee 
		where $H_n$ denotes the last three terms of \eqref{heff_spinful_app}, sums over lattice sites and spin indices have been left implicit, and 
		\be \wt \mca^a_{r,r'} = |t_d|(\d_{r,r'+a} + \d_{r,r'-a}).\ee 
		The functional integral over $D^s$ is however only well-defined if $st_d>0$. If we are in a situation where $st_d<0$, we must first  perform a change of variables which shifts the momenta of $D^{a\pm}$ by $\pi$ in the $a$ direction: 
		\be D_{r\s}^{as} \mt i (-1)^{r_a} D_{r\s}^{as}, \qq d_{r\s}^{as} \mt (-1)^{r_a} d_{r\s}^{as}.\ee 
		In terms of the fermion fields, this is affected by taking $f_{r\s} \mt e^{\pi i \sum_a r_a^2/2}f_{r\s}$.\footnote{Strictly speaking this doesn't actually do anything as $\mca^a$ has equally many positive and negative eigenvalues; however, we will restrict the integral over $D^{as}$ to one over slowly-varying fields (after the possible shift above), for which the eigenvalues of $\mca^a$ are positive, and the sign of the quadratic term for $D^{as}$ is meaningful. } In the tilted lattice parametrization with attractive bare $V_0$ we have $t_d<0$---thus in this case the $SV(2)$-preserving dipoles $d^{a+}_{r\s}$  will condense at momentum $\pi$.\footnote{We could also consider having $V_0>0$ but $V < 0$ (so that $V<0$ but $t_d  >0$), but this necessitates having $4t_d > V_0 \implies t/ \De> 1/2$, which is rather unrealistic.}
		
		After performing this shift if needed, we now integrate out the fermions. Analogously to the spinless calculation, this gives an action of the form \eqref{eft} with coupling constants 
		\bea \kappa_D^{+} & = \frac{1}2\sum_{\l=\pm1} \frac1{(\sfV + (2\l-1)J/4)^3} \\ 
		\kappa_D^{-} & = \frac1{(\sfV+ J/4)^3} \\ 
		K^{a+}_b & = K^{a-}_b = \frac14 \d_{a,b} \\ 
		r^+ & = \frac1{2|t_d|} - \frac12 \sum_{\l=\pm1}  \frac1{\sfV+ (2\l-1)J/4} \\ 
		r^- & = \frac1{2|t_d|} - \frac1{\sfV+ J/4}.
		\eea 
		If we allow ourselves to tune $t_d, \sfV, J$ independently, there are clearly regions in parameter space where the $r^\pm$ are negative. What about when the parameters are fixed in the way that they are in the tilted lattice parametrization? For simplicity, we will take the ratio $t_{a,\s}^2 / \De_{a,\s}^2 \equiv \sft^2$ of the single particle hopping to tilt strength to be independent of both $a$ and $\s$. Since we have restricted ourselves to a situation with attractive $V = V_0 - 4t_d < 0$ (with $t_d = \eta V_0$), we must have either $V_0 < 0, \, \sft < 1/2$, or $V_0 > 0,\, \sft > 1/2$; since we are doing perturbation theory in $\sft$ the former is more natural, and we will restrict our attention to this case. We then find 
		\bea r^+ & =  \frac{\sft^2}{2|V_0|} \( 1- \frac1{1/\sft^2 - (4d+2)} - \frac1{1/\sft^2 - (4d+6)}\)  \\  
		r^- & = \frac{\sft^2}{|V_0|} \(\frac12 - \frac1{1/\sft^2 - (4d+2)}\).  \eea 
		Then the requirement that the $r^\pm$ be negative translates into the requirement that  
		\bea \frac1{\sqrt{4d+5+\sqrt5}} <\,  &\,\,\sft^+ < \frac1{\sqrt{4d+6}}\\
		\frac1{\sqrt{4d+4}} < &\,\, \sft^- <\,  \frac1{\sqrt{4d+2}} \eea 
		This gives a (rather narrow) window in which a dipole condensate can form, with $r^+$ going negative first as $\sft$ is increased. 
		
		\sss{Replusive $V$}
		
		We now consider the case of repulsive $V$. In the tilted lattice context, repulsive bare interactions give a {\it ferro}magnetic spin exchange term (by way of \eqref{spinful_coeffs}), and it is thus natural to start the mean-field analysis from the state $\k{FERRO} = \k{\upa}^L$. Since this state has maximal $S^z_{tot}$, $d^+$ dipoles cannot be created on top of it, although it is still possible to consider a transition where the spin-1 $d^-$ dipoles condense and break $SU(2)$.\footnote{If one considers $J>0$ instead then the starting point is an antiferromagnet, on top of which $d^+$ dipoles can condense. An analysis of this case can be done similarly, but for brevity we will focus on the case relevant for the tilted lattice setup. }
		
		The effective quadratic action for the dipole fields $D^{a-}_{r\s}$ is the same as in the case of attractive $V$, except with $\k{CDW}$ replaced by $\k{FERRO}$. The states created by $d^{a-}_{r\doa}$ and $d^{a-\da}_{r\upa}$ when acting on $\k{FERRO}$ are eigenstates of $H_n$ with relative energy 
		\be E_{n0} = V - V' - J(d-1/4).\ee 
		Because of the ferromagnetic exchange a mean-field solution is clearly hopeless at large $d$---the increase in spin exchange energy is too great to be overcome by the reduction in dipole kinetic energy. Indeed, the mass of the $D^{a-}_{r\s}$ fields is 
		\be r^- = \frac1{2|t_d|} - \frac{1}{V-V'-J(d-1/4)} = \frac{\sft^2}{V_0} \( \frac12 - \frac1{1/\sft^2 + 4d-6}\) ,\ee 
		which in the tilted lattice parametrization is always positive if $d>1$, and at $d=1$ is negative only if $\sft > 1/2$ (which is a bit too large to be comfortable within perturbation theory). For these reasons a DC forming in the repulsive case appears to be rather unlikely without further help from other terms in the Hamiltonian (e.g. non-onsite Hubbard interactions). 
		
		\section{Solution of the 1d spinless dipolar Fermi-Hubbard model} \label{app:spinless1d}
		
		In this appendix we briefly discuss the 1d spinless DFHM, which admits a simple solution in terms of a spin half model that maps onto an XXZ chain. The Hamiltonian is 
		\bea \label{1ddfhm} H & = -\sum_{i} t_{d} (d^{\da}_{i}d^a_{i+2} + h.c.)  + \sum_j \(V n_j n_{j+1} +  V' n_j n_{j+2}\),\eea 
		where in the tilted lattice parametrization with tilt $\De$, bare hopping $t$, and bare nearest-neighbor repulsion $V_0$, we have $t_d = t^2V_0 / \De^2, \, V = V_0(1-2t^2/\De^2), \, V' = 2V_0t^2 / \De^2$. 
		
		As explained in \cite{lake2022dipole} (see also \cite{moudgalya2019thermalization} for a discussion of a closely related model), the natural picture of ground states of \eqref{1ddfhm} is that they are obtained from `resonating' the state $|0110011\dots\ran$. Letting $\mcp$ denote the projector onto the Krylov subspace of this state, we find that $\mcp H \mcp$ becomes a spin model on the effective spin-1/2 Hilbert space with $\k\upa \equiv |0_{2i}1_{2i+1}\ran, \, \, \k\doa \equiv \k{1_{2i}0_{2i+1}}$ (it is easy to prove that $\k{0_{2i}0_{2i+1}}$ and $\k{1_{2i}1_{2i+1}}$ never appear in this Krylov sector, at least if we only keep the 4-site dipole hopping term as in \eqref{1ddfhm}). Indeed, a simple calculation shows that the projected version of \eqref{1ddfhm} becomes an XXZ model: 
		\be \mcp H \mcp  = \frac{t_d}2 \sum_i (X_i X_{i+1} + Y_i Y_{i+1}) - \frac{V-2V'}4 \sum_i Z_i Z_{i+1}.\ee 
		Within this approximation we thus expect a transition between a gapped CDW and a gapless state at $t_d \approx \pm (V-2V') / 2$. 
		In the tilted lattice parametrization, this occurs when 
		\be \label{1destimate} t_d = \frac{V_0-6t_d}2 \implies t/\De = \frac1{\sqrt 8}.\ee 
		The factor of $1/\sqrt8\approx 0.35$ is small enough that we can contemplate a scenario in which perturbation theory remains valid in this regime, meaning that the transition can indeed be accessed in the tilted lattice setup. Fig. \ref{fig:1ddfhm} shows results from DMRG run on the fermion Hamiltonian \eqref{1ddfhm}, which confirm that the transition indeed happens near $t_1/ \De = 1/\sqrt 8 \approx 0.35$. 
		
		\begin{figure}
			\includegraphics[width=.33\tw]{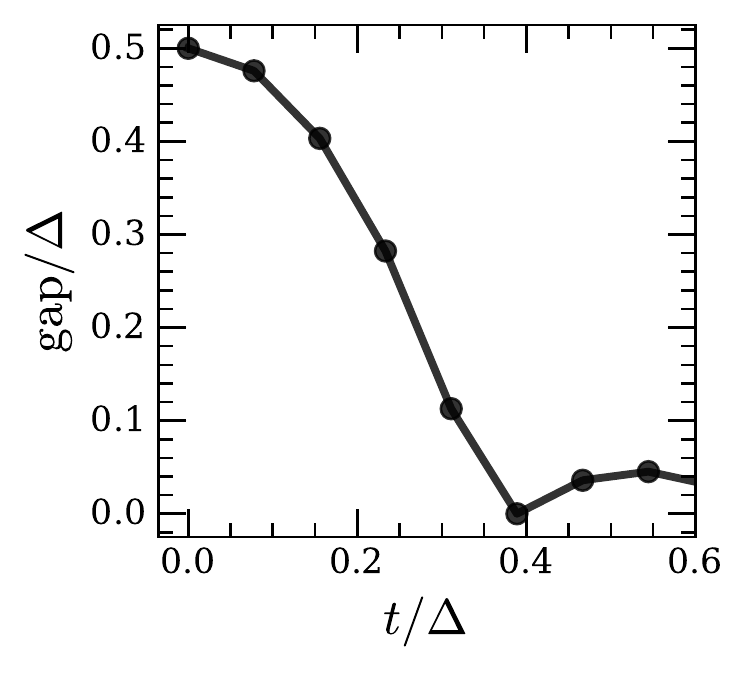}
			\caption{\label{fig:1ddfhm} The energy gap of the 1d DFHM \eqref{1ddfhm} obtained from DMRG with bond dimension $\chi = 150$ on a $L=40$ ring. We take the couplings in \eqref{1ddfhm} to be parametrized in terms of the single-particle hopping $t$ and tilt strength $\Delta$. The spin-1/2 mapping predicts a gap closing at $t/\De = 1/\sqrt{8}\approx 0.35$. At $t/\De$ below the transition we find a period-2 CDW with short-ranged dipole-dipole correlation functions, while above the transition the density is uniform, with dipole correlators decaying as power laws. }
			
		\end{figure}

	\end{widetext}

\end{document}